\documentclass[twocolumn,showpacs,preprintnumbers,amsmath,amssymb,prl]{revtex4}

\usepackage{graphicx}
\usepackage{bm}

\begin{document}

\title{Silica-Like Sequence of Anomalies in Core-Softened Systems}

\author{Yu. D. Fomin}
\affiliation{Institute for High Pressure Physics, Russian Academy
of Sciences, Troitsk 142190, Moscow Region, Russia}

\author{E. N. Tsiok}
\affiliation{Institute for High Pressure Physics, Russian Academy
of Sciences, Troitsk 142190, Moscow Region, Russia}

\author{V. N. Ryzhov}
\affiliation{Institute for High Pressure Physics, Russian Academy
of Sciences, Troitsk 142190, Moscow Region, Russia}
\affiliation{Moscow Institute of Physics and Technology, 141700
Moscow, Russia}

\date{\today}

\begin{abstract}
In this paper we present a simulation study of density, structural
and diffusion anomalies in core-softened system introduced in our
previous publications. It is well-known, that with appropriate
parametrization, core-softened systems are remarkable model
liquids that exhibit anomalous properties observed in tetrahedral
liquids such as silica and water. It is widely believed that
core-softened potentials demonstrate the water-like sequence of
anomalies. We show that with increasing the depth of the
attractive part of the potential the order of the region of
anomalous diffusion and the regions of density and structural
anomalies is inverted and have the silica-like sequence. We also
show that the slope of the Widom line is negative like in water.
\end{abstract}

 \maketitle


In the last decades there is a growing interest to a class of
fluids that display anomalous thermodynamic and kinetic behaviors.
The most common and well known example is water. The water phase
diagram has regions where a thermal expansion coefficient is
negative (density anomaly), self-diffusivity increases upon
compression (diffusion anomaly), and the structural order of the
system decreases with increasing pressure (structural anomaly)
\cite{ad1,stanley1,deben2001,netz,book,book1}. Later on it was
discovered that many other substances also demonstrate similar
behavior. Some typical examples are silica, silicon, phosphorus
and many others \cite{book,book1,si2,br,p1,p2}.

As it was found in experiments \cite{ad1} and simulations
\cite{deben2001,netz}, the water anomalies have a well-defined
sequence: the regions where these anomalies take place form nested
domains in the density-temperature \cite{deben2001} (or
pressure-temperature \cite{netz}) planes: the density anomaly
region is inside the diffusion anomaly domain, and both of these
anomalous regions are inside a broader structurally anomalous
region. This water-like behavior was found in systems with
spherically symmetric core-softening potentials with two length
scales
\cite{fr1,buld2009,fr2,fr3,fr4,errington,20,yan2005,yan2006,wejcp,wejcp1,wepre,wepre1,wepla,we_cm,we_inv,we2013}.

However, in other anomalous systems the sequence of anomalies may
be different. For example, in computer simulation of the system
with the Van Beest-Kramer-Van Santen (BKS) potential the hierarchy
of anomalies for silica is different compared to water \cite{si2}.
In this case the diffusion anomaly region contains the structural
anomalous region which, in turn, has the density anomaly region
inside. To our knowledge this is the only example of such
inversion of the order of the anomalies discussed in literature
till now.

It is widely believed that in the core-softened systems hierarchy
of anomalies has the water-like type. For example, Yan {\it et
al.} \cite{yan2005,yan2006} characterized the structural,
thermodynamic, and kinetic properties of a family of discontinuous
core-softened potentials that vary in the length scale of the soft
repulsion region. In each case, they found the same relationship
between anomalous regions as observed for water: structural
anomalies preceded diffusivity anomalies, which preceded density
anomalies. In Ref. \cite{errington} it was discussed an approach
for the analysis of the anomalous behavior based on the well known
Rosenfeld scaling relations \cite{ros} which connect the transport
coefficients with excess entropy. In this case there was obtained
the explicit equation determining the appearance of the anomalies
in system \cite{errington}:
\begin{equation}
\left(\frac{\partial S_{ex}}{\partial \ln \rho}\right)_T>c,
\label{cond}
\end{equation}
where $\rho$ is the number density, $S_{ex}=S-S_{id}$ is the
excess entropy, equal to difference between total $S$ and ideal
gas $S_{id}$ entropies, and $c$ is the property-specific constant.
Based on Rosenfeld's scaling parameters \cite{errington}, it was
shown that value $c=0$ corresponds to structural anomaly, $c=0.42$
to diffusion anomaly, and $c=1$ - to density anomaly. From
(\ref{cond}) one can conclude that anomalous behavior always occur
in the water-like order: structural anomalies precede diffusivity
anomalies, which in turn precede density anomalies. This
conclusion was verified by the computer simulation of the system
interacting through a two-scale potential introduced by Jagla
\cite{jagla}. However, as it was shown in our previous
publications \cite{wepre1,we_cm}, Rosenfeld scaling fails in the
vicinity of anomalies,  and it can not be used for the correct
analysis of the order of the anomalies.

It is important to note that the equation (\ref{cond}) contains
two conditions which are basing on strict thermodynamic arguments:
the conditions for structural and density anomalies
\cite{errington}. This means that the density anomaly should {\it
always} follow after the structural anomaly. On the other hand,
the diffusion anomaly can be located at every place. For example,
for silica the diffusion anomaly precedes the structural and
density anomalies.

This paper presents a simulation study of anomalies in
core-softened system introduced in our previous publications
\cite{wejcp,wejcp1,wepre,wepre1,wepla,we_cm,we_inv,we2013}. We
investigate the sequence of the anomalous regions and find that
with increasing the attractive part of the potential the system
demonstrates both water-like and silica-like behavior. It is also
shown that the potential qualitatively correctly reproduces the
behavior of the Widom line for the liquid-liquid transition.

In the present study we investigate a system of particles
interacting via the potential with "hard" core, repulsive shoulder
and attractive well \cite{wejcp1,we_cm}.

The general form of the potential is written as
\begin{eqnarray}
U(r)&=&\varepsilon\left(\frac{\sigma}{r}\right)^{14}+\varepsilon\left(\lambda_0-
\lambda_1\tanh(k_1\{r-\sigma_1\})\right.+\nonumber\\
&+&\left.\lambda_2 \tanh(k_2\{r-\sigma_2\})\right). \label{3}
\end{eqnarray}

\begin{figure}
\includegraphics[width=8cm]{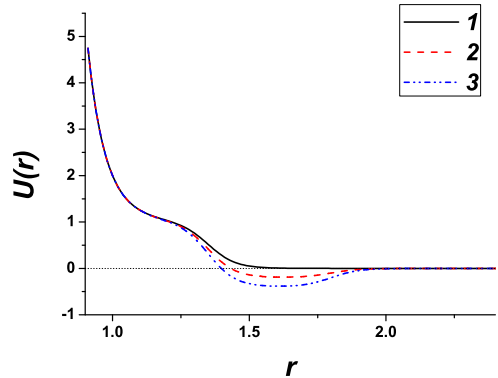}%

\caption{\label{fig:fig02} (Color online) Family of the potentials
with $\sigma_1=1.35$ and different attractive wells. The curves
are numerated in accordance with Table 1.}
\end{figure}

Here $k_1=k_2=10.0$ and the parameters of the potentials (\ref{3})
are given in Table 1. The family of the potentials with
$\sigma_1=1.35$ and different attractive wells is shown in
Fig.~\ref{fig:fig02}. It should be noted that the potential
(\ref{3}) is very similar to the Fermi-Jagla potential suggested
recently in Ref. \cite{buld2011}, where the Fermi distribution
function is used except for the hyperbolic tangent in Eq.
(\ref{3}) in order to describe the smoothed step \cite{pot_fermi}.

\begin{table}
\begin{tabular}{|c|c|c|c|c|c|c|}
  \hline
  number & $\sigma_1$ & $\sigma_2$& $\lambda_0$  & $\lambda_1$ & $\lambda_2$ & well depth \\
  \hline
  1 & 1.35 & 0 & 0.5 & 0.5 & 0 & 0\\
  2 & 1.35 & 1.80 & 0.5 & 0.60 & 0.10 & 0.20\\
  3 & 1.35 & 1.80 & 0.5 & 0.7 & 0.20 & 0.4\\
  \hline

\end{tabular}

\caption{The potential parameters used in simulations (Eq.
(\ref{3})).}

\end{table}

In the remainder of this paper we use the dimensionless
quantities: $\tilde{{\bf r}}\equiv {\bf r}/\sigma$,
$\tilde{P}\equiv P \sigma^{3}/\varepsilon ,$ $\tilde{V}\equiv V/N
\sigma^{3}\equiv 1/\tilde{\rho}, \tilde{T} \equiv k_BT/\varepsilon
$. As we will only use these reduced variables, we omit the
tildes.

In Refs.
\cite{wejcp,wejcp1,wepre,wepre1,wepla,we_cm,we_inv,we2013} it was
shown that these systems demonstrate anomalous behavior. A
relation between phase diagram and anomalous regions was also
discussed in these articles. Our later publications gave detailed
study of diffusion, density and structural anomalies in this
system \cite{wepre1,wepla,we_cm,we_inv,we2013}.

In the present article we carry out a molecular dynamics and Monte
Carlo study of the core-softened systems  and monitor the change
in phase diagram and anomalous regions with changing the potential
parameters. The details of simulations can be found in Ref.
\cite{we_inv}.

In order to find the melting lines we carry out the free energy
calculations for different phases and construct a common tangent
to them. For our potentials we computed the free energy of the
liquid by integrating the equation of state along an isotherm
\cite{book_fs}:
$\frac{F(\rho)-F_{id}(\rho)}{Nk_BT}=\frac{1}{k_BT}\int_{0}^{\rho}\frac{P(\rho')-\rho'
k_BT}{\rho'^2}d\rho'$. Free energies of different crystal phases
were determined by the Monte Carlo simulations with the method of
coupling to the Einstein crystal \cite{book_fs}. In this case the
excess entropy can be computed via $S_{ex}=\frac{U-F_{ex}}{N
k_BT}$, where $U$ is the internal energy \cite{book_fs}. The total
entropy is $S=S_{ex}+S_{id}$, where the ideal gas entropy is
$\frac{S_{id}}{Nk_B}=\frac{3}{2}\ln(T)-\ln(\rho)+\ln(\frac{(2 \pi
mk_B)^{3/2}e^{5/2}}{h^3})$.

Here we present the anomalous regions for the three systems shown
in Fig.~\ref{fig:fig02}.

The density anomaly means that density increases upon heating or
that the thermal expansion coefficient becomes negative. Using the
thermodynamic relation $\left(\partial P/\partial
T\right)_V=\alpha_P/K_T$, where $\alpha_P$ is a thermal expansion
coefficient and $K_T$ is the isothermal compressibility and taking
into account that $K_T$ is always positive and finite for systems
in equilibrium not at a critical point, we conclude that density
anomaly corresponds to minimum of the pressure dependence on
temperature along an isochor. This is the most convenient
indicator of density anomaly in computer simulation.

Initially the structural anomaly was introduced via order
parameters characterizing the local order in liquid
\cite{deben2001,si2,yan2005,20,top,snr,top1}. However, later on
the local order was also related to excess entropy of the liquid
which is defined as the difference between the entropy and the
ideal gas entropy at the same $(\rho,T)$ point: $S_{ex}=S-S_{id}$.
In normal liquid excess entropy is monotonically decaying function
of density along an isotherm while in anomalous liquids it
demonstrates increasing in some region. This allows to define the
boundaries of structural anomaly at given temperature as minimum
and maximum of excess entropy.

\begin{figure}
\includegraphics[width=8cm]{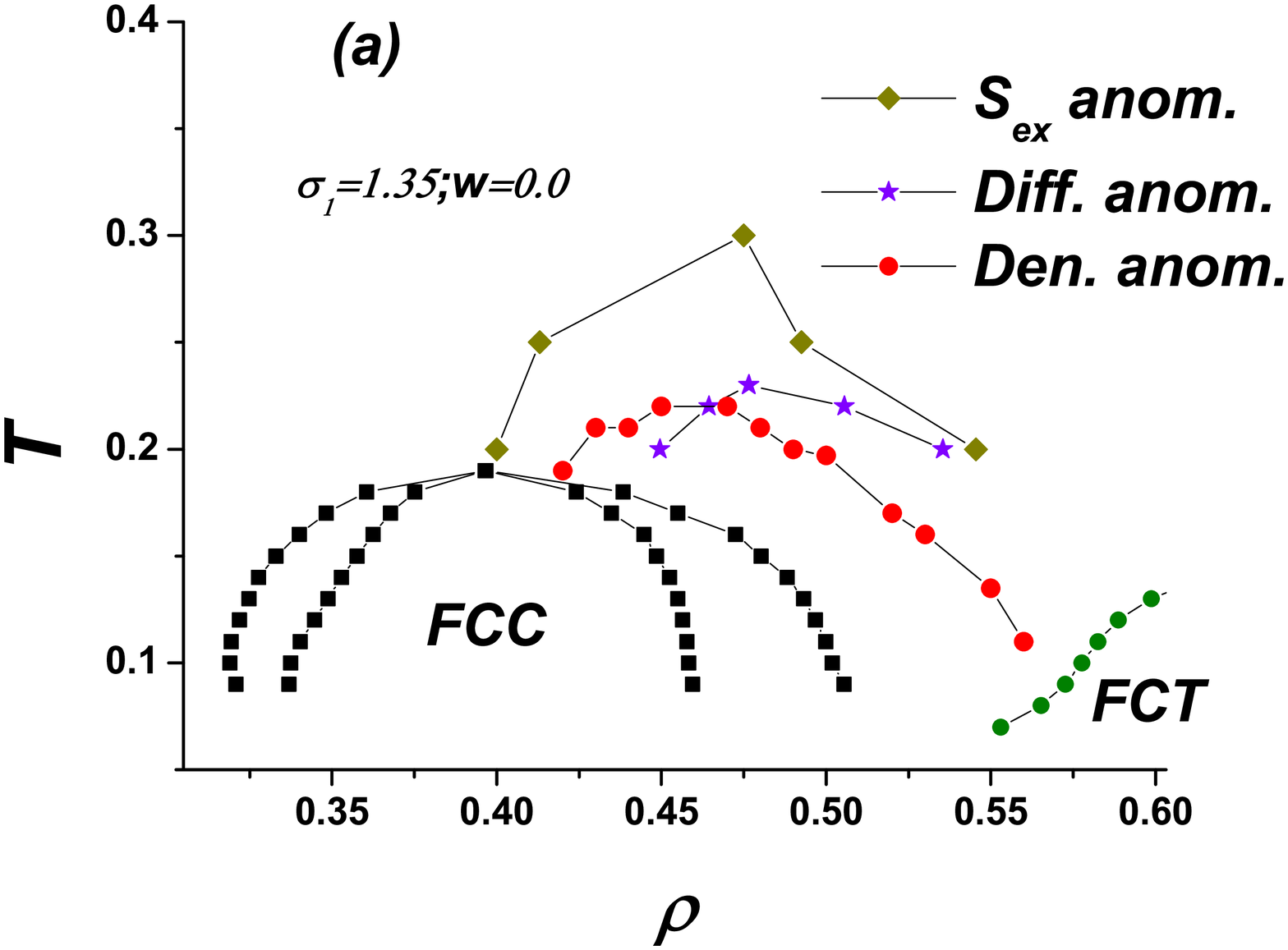}

\includegraphics[width=8cm]{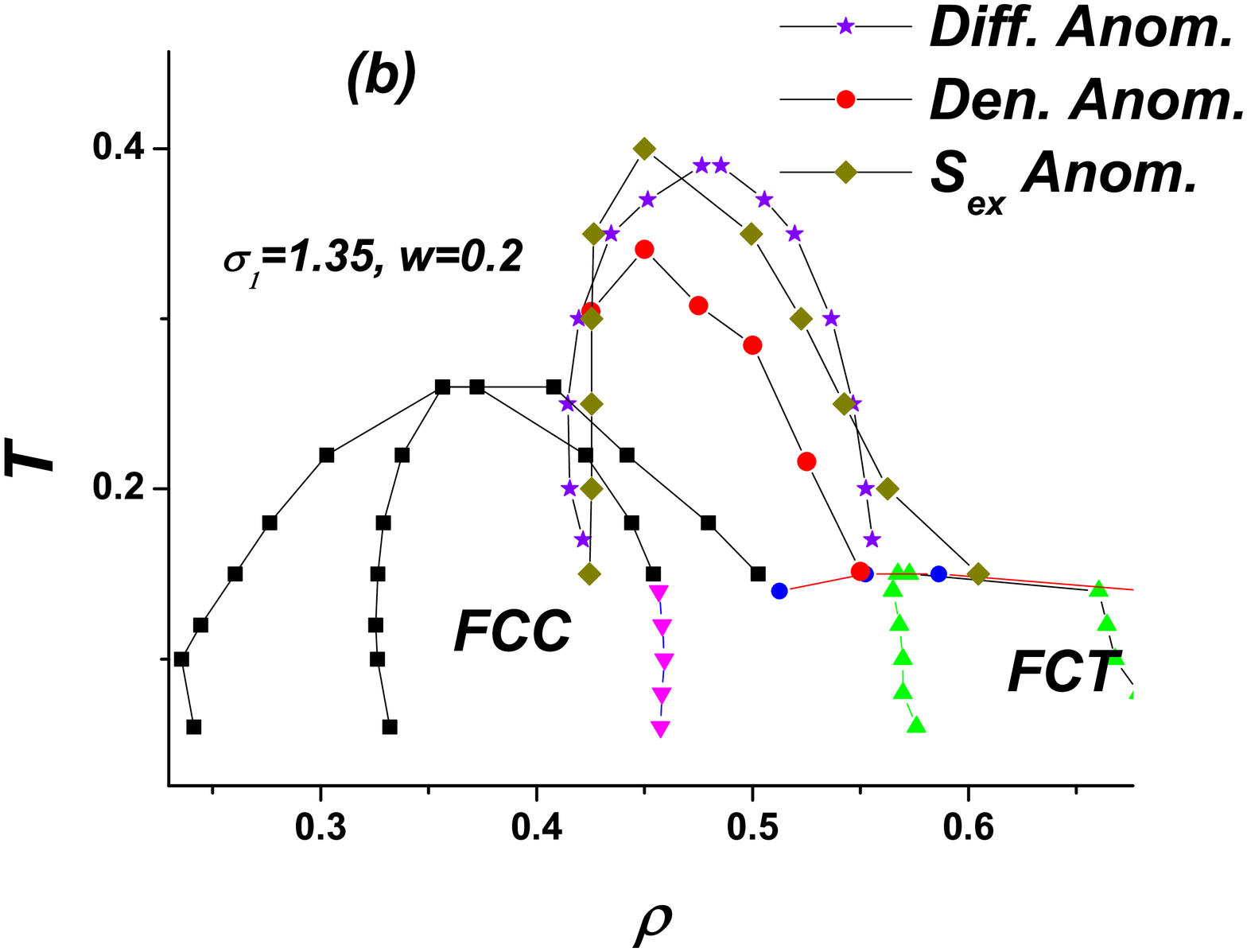}

\includegraphics[width=8cm]{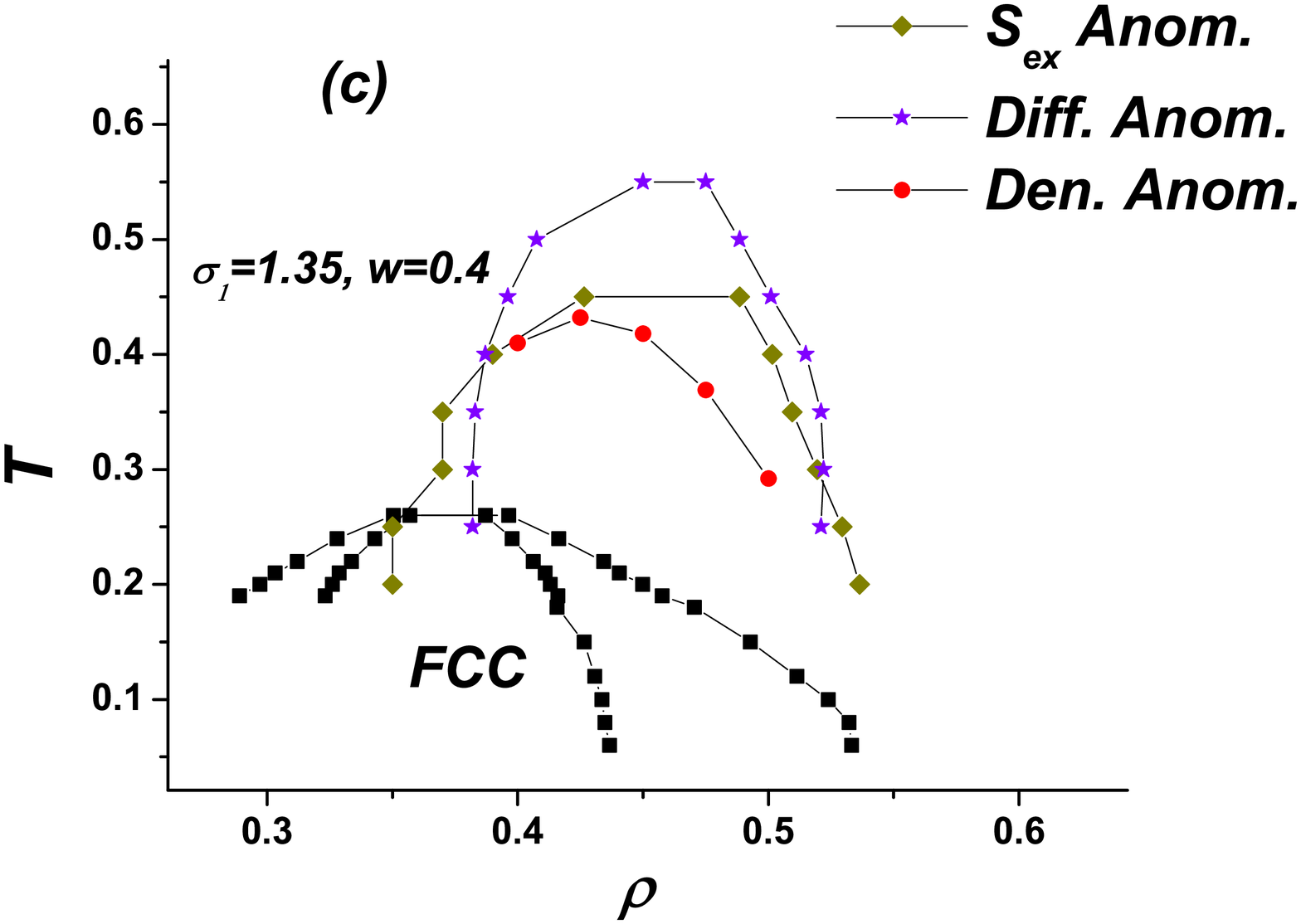}

\caption{\label{fig:fig2} (Color online) (a) Location of anomalous
regions at the low density part of the phase diagram of the system
with $\sigma_1=1.35$, where Face Centered Cubic (FCC)  and Face
Centered Tetragonal (FCT) phases are shown. The water-like order
of anomalies takes place: the density anomaly region is inside the
diffusion anomaly domain, and both of these anomalous regions are
inside a broader structurally anomalous region. Location of
anomalous regions at the phase diagram for (b) system with
$\sigma_1=1.35$ and $w=0.2$; (c) $\sigma_1=1.35$ and $w=0.4$ with
the silica-like order of anomalies (see Table 1).}
\end{figure}

The behavior of the diffusion coefficient, pressure and excess
entropy for the purely repulsive system with $\sigma_1=1.35$ has
been discussed, for example, in Refs.
\cite{wejcp,wepre1,we_cm,we_inv,we2013}. One can see that all
three anomalies take place in the system. Fig.~\ref{fig:fig2}(a)
places the regions of the anomalies in the phase diagram. As it
was shown in \cite{wejcp,wejcp1,we_inv} (see, for example, Fig. 1
in Ref. \cite{wejcp1}), the phase diagram of the system consists
of the high density and low density Face Centered Cubic (FCC)
phases, corresponding to hard core and repulsive shoulder parts of
the potential, separated by a sequence of crystalline phases. In
Fig.~\ref{fig:fig2}(a) we show the low density part of the phase
diagram with the FCC and FCT (Face Centered Tetragonal) phases.

One can see that the anomalous regions correspond to the picture
proposed for water \cite{deben2001}, i.e. the diffusion anomaly
region is inside the structural anomaly and the density anomaly is
mainly inside the diffusion anomaly.

In Ref. \cite{we_inv} it was shown that in the system with the
purely repulsive potential the diffusion and density anomalies
inverted with respect to each other with increasing the repulsive
core diameter, i.e. now diffusion anomaly region is inside the
density anomaly one.

Next we consider the influence of attraction on the anomalous
behavior of the system. For this, we study the system with step
size $\sigma_1=1.35$ and different well depths (Table 1 and
Fig.~\ref{fig:fig02}). In this case one can see that with
increasing the depth of the attractive well the sequence of the
anomalies also inverted: for $w=0$ we have the water-like order of
the anomalies (see Fig.~\ref{fig:fig2}(a)), at $w=0.2$ the
locations of the diffusion and structural anomalies almost
coincide (Fig.~\ref{fig:fig2}(b)). At last, for $w=0.4$ the
diffusion anomaly region contains the structural anomalous region
which, in turn, has the density anomaly region inside
(Fig.~\ref{fig:fig2}(c)) \cite{remark}. As a result, for $w=0.4$
we obtain the configuration of anomalies which is the same as in
silica \cite{si2}.

\begin{figure}
\includegraphics[width=8cm]{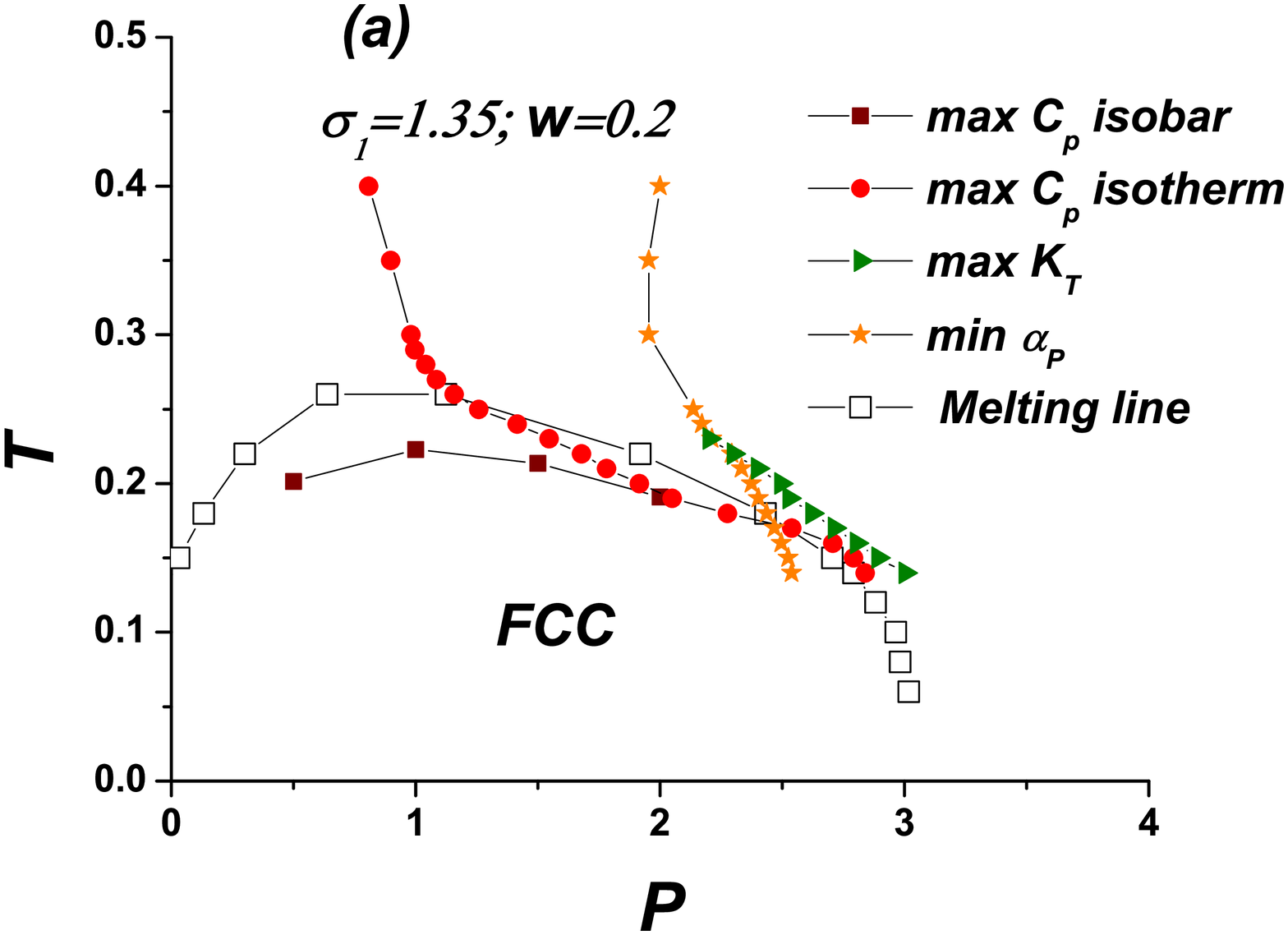}

\includegraphics[width=8cm]{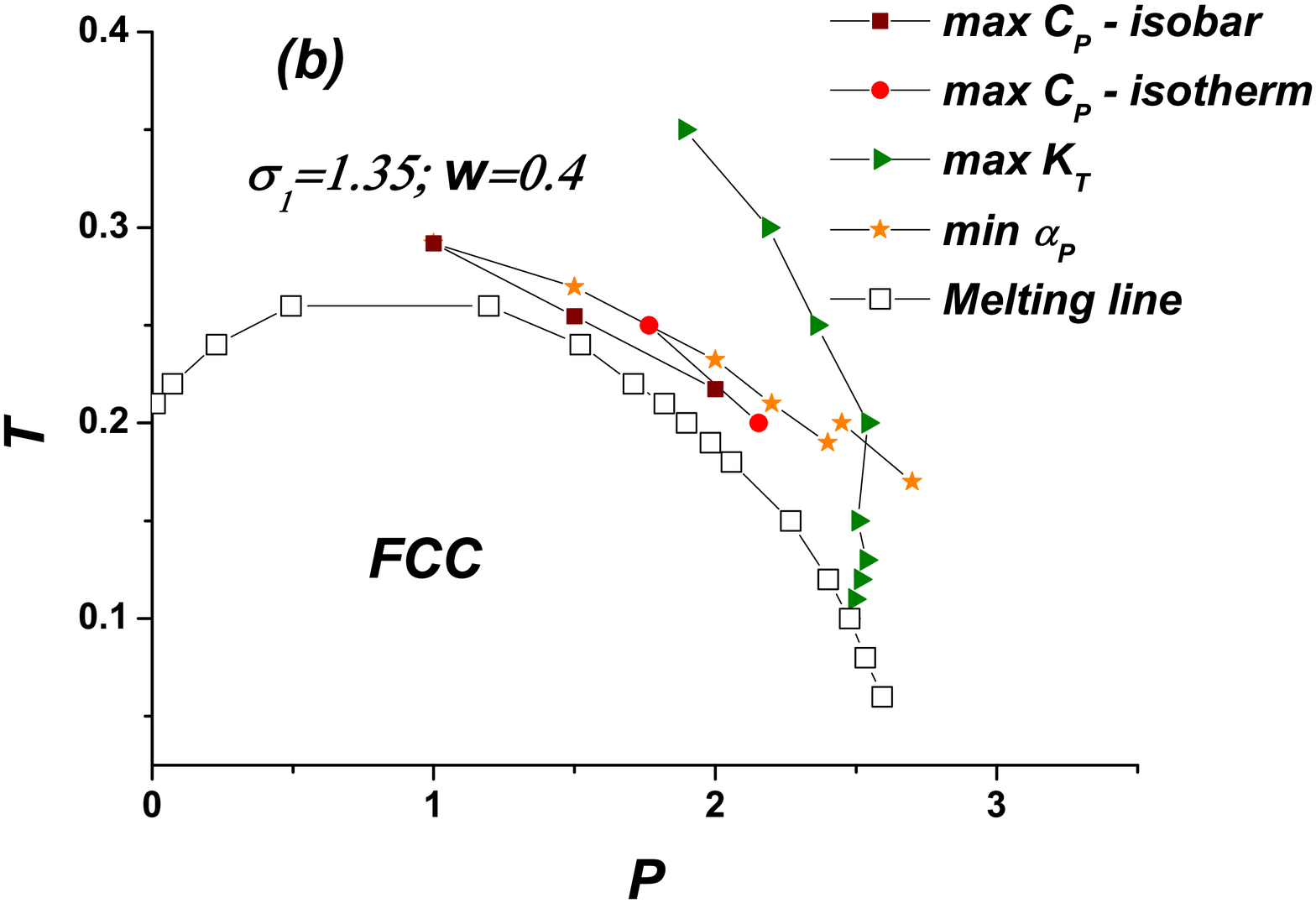}

\caption{\label{fig:fig22} (Color online) The maxima lines of
isobaric heat capacity $C_P$ along isobars and isotherms, the line
of the isothermal compressibility maxima $K_T$, the line of the
thermal expansion coefficient minima along with the melting line
for the systems 2 (a) and 3 (b) (Table 1).}
\end{figure}

It seems that a most popular point of view is that the
hypothesized liquid-liquid critical point is the thermodynamic
source of all water anomalies \cite{PNAS1,poole,mish,fran}, the
terminal point of a line of first-order liquid-liquid phase
transition. The line emanating from this critical point is
sometimes called the Widom line and is often considered as an
extension of the coexistence line into the one-phase region (see,
for example, \cite{PNAS1,buld2011,BR2011}).This line is determined
by the lines of the maxima of the thermodynamic response functions
which asymptotically approach one another as the critical point is
approached \cite{PNAS1,BR2011}. The lines of the liquid-liquid
phase transition in the computer simulations of water \cite{PNAS1}
and silicon \cite{sast2} have the negative slope. On the other
hand, earlier computer simulations of the isotropic core-softened
potentials (see, for example, \cite{PNAS1}) suggested that the
slope of the Widom line is positive in these systems.

In Fig.~(\ref{fig:fig22}) we show the maxima lines of isobaric
heat capacity $C_P$ along isobars and isotherms, the line of the
isothermal compressibility maxima $K_T$, the line of the thermal
expansion coefficient $\alpha_P$ minima along with the melting
line for the systems 2 and 3 (Table 1). One can see, that for our
potential the Widom line has a slope which coincides with the
results for water \cite{PNAS1} and silicon \cite{sast2}. It should
be noted that the similar results were obtained recently
\cite{buld2011} for the potential similar to (\ref{3}).

\begin{figure}
\includegraphics[width=8cm]{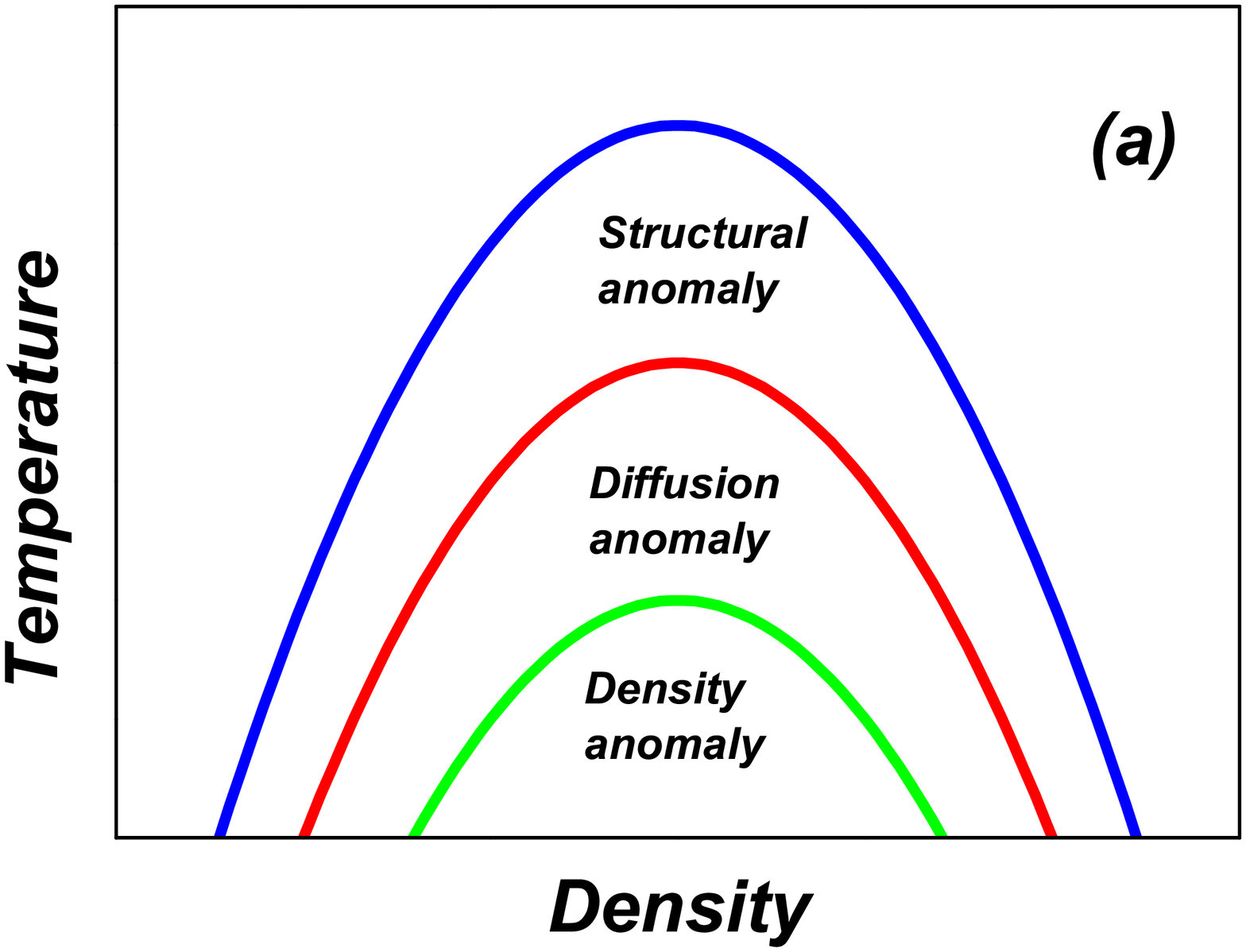}

\includegraphics[width=8cm]{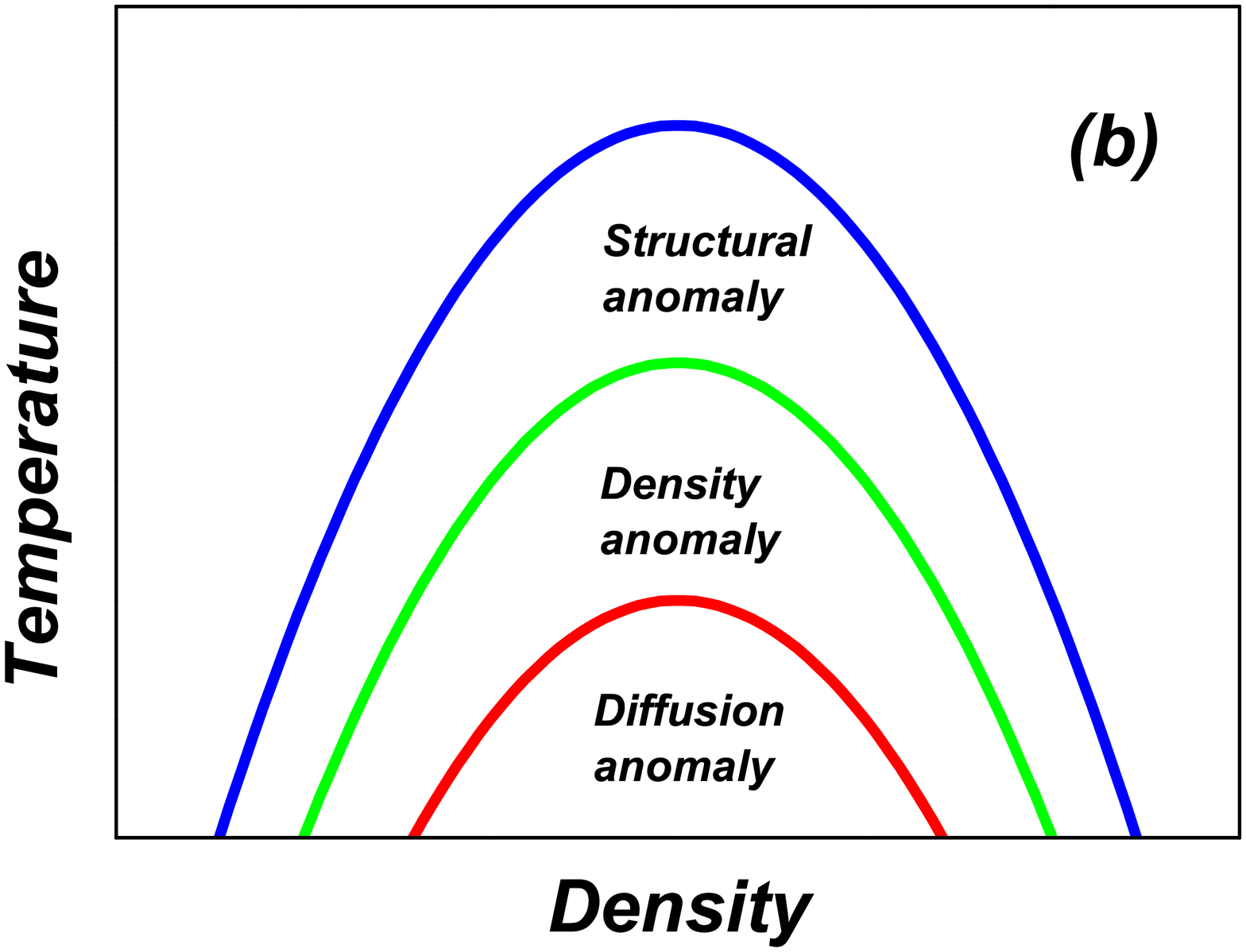}

\includegraphics[width=8cm]{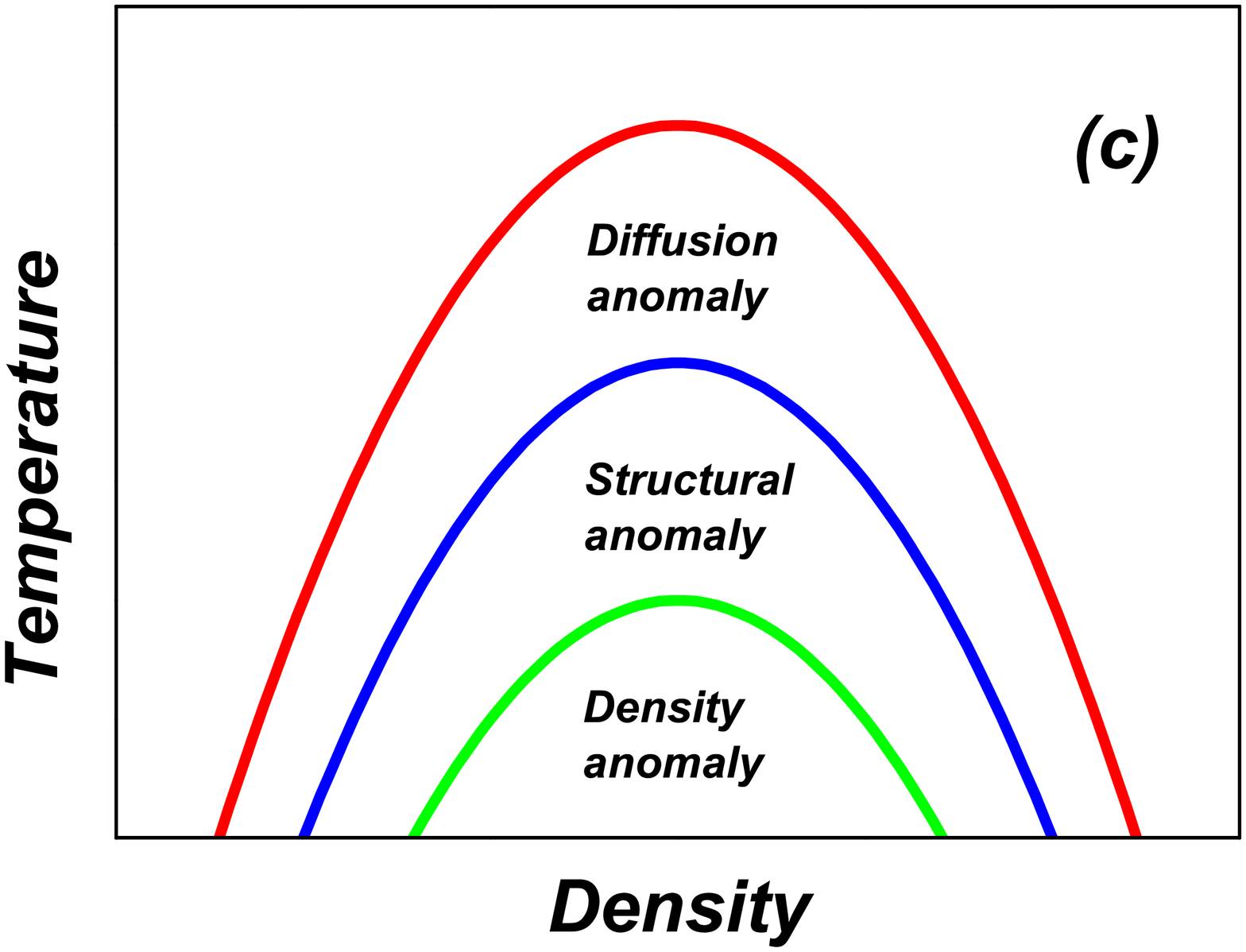}

\caption{\label{fig:fig3} (Color online) Schematic of the regions
wherein structural, diffusivity, and density anomalies are found
within the temperature-density planes. Region of anomalous density
behavior always appears as nested dome within the structural
anomalous envelope while the diffusivity anomaly may be (a)
between the structural and density ones (see Fig.~\ref{fig:fig2}
(water-like behavior); (b) inside the density anomalous region;
and (c) the outermost envelope (silica-like behavior).}
\end{figure}

In conclusion, this publication represents a detailed computer
simulation study of anomalous behavior of core-softened systems
proposed in our publications
\cite{wejcp,wejcp1,wepre,wepre1,wepla,we_cm,we_inv,we2013}. The
core-softened potentials are widely investigated because they
reproduce the water-like anomalies. Taking into account that the
anomalies also exist in the systems where the hydrogen bonds are
absent, it seems that the unusual properties of water are quite
universal and can be investigated with the help of the isotropic
core-softened potentials. To our knowledge, all isotropic
core-softened potentials which were considered in the previous
publications show the sequence of anomalies characteristic for
water: the density anomaly region is inside the diffusion anomaly
domain, and both of these anomalous regions are inside a broader
structurally anomalous region. On the other hand, in other
anomalous systems the sequence of anomalies may be different. For
example, the hierarchy of anomalies for silica is different
compared to water. In the present article we analyze the
possibility of changing the order of anomalies regions depending
on the parameters of the potential. It is shown that for the
potential (\ref{3}) for small values of the repulsive step the
sequence of anomalies is the same as in water, however, with
increasing the width of the repulsive shoulder the order of the
region of anomalous diffusion and the region of density anomaly is
inverted. With increasing the depth of the attractive well we
obtain the configuration of anomalies which is the same as in
silica \cite{si2}, where the diffusion anomaly region contains the
structural anomalous region which, in turn, has the density
anomaly region inside. It seems that this is the first case when
the isotropic core-softened system demonstrates the sequence of
anomalies which is different from the water one and may be the
same as in silica. It is also shown that for the potential
(\ref{3}) the slope of the Widom line is negative as in the case
of water and silicon.


We thank S. M. Stishov and V. V. Brazhkin for stimulating
discussions. Y.F. and E.T. also thanks Russian Scientific Center
Kurchatov Institute and Joint Supercomputing Center of Russian
Academy of Science for computational facilities. The work was
supported in part by the Russian Foundation for Basic Research
(Grants No 11-02-00341-a, 13-02-00579-a, and 13-02-00913-a)  and
the Ministry of Education and Science of Russian Federation
projects 8370 and 8512.


\begin{thebibliography}{99}
\bibitem{remark} We would like to mention, that in our previous paper \cite{we_cm}
there is an error in Fig.17(c), and it should be replaced by the
present Fig.~\ref{fig:fig2}(c).
\end{thebibliography}
\end{document}